\documentclass[aps,prl,preprint,showpacs,superscriptaddress,groupedaddress]{revtex4}%

\usepackage{graphicx}
\usepackage{dcolumn}
\usepackage{bm}
\usepackage[caption = false]{subfig}
\usepackage{lipsum}
\begin{document}
\title{Magnetization Dynamics of Topological Defects and the Spin Solid in Kagome Artificial Spin Ice}
\author{V.S.~Bhat} \email[]{vinayak.bhat@epfl.ch} \affiliation{Lehrstuhl f\"{u}r Physik funktionaler Schichtsysteme, Physik Department E10, Technische Universit\"{a}t M\"{u}nchen, 85748 Garching, Germany}\affiliation{\'Ecole Polytechnique F\'ed\'erale de Lausanne, Sciences et Techniques de l'Ing\'enieur, Institut des Mat\'eriaux, Laboratoire des Mat\'eriaux Magn\'etiques Nanostructur\'es et Magoniques, 1015 Lausanne, Switzerland}

\author{F.~Heimbach}\affiliation{Lehrstuhl f\"{u}r Physik funktionaler Schichtsysteme, Physik Department E10, Technische Universit\"{a}t M\"{u}nchen, 85748 Garching, Germany}
\author{I.~Stasinopoulos} \affiliation{Lehrstuhl f\"{u}r Physik funktionaler Schichtsysteme, Physik Department E10, Technische Universit\"{a}t M\"{u}nchen, 85748 Garching, Germany}
\author{D.~Grundler} \email[]{dirk.grundler@epfl.ch}  \affiliation{\'Ecole Polytechnique F\'ed\'erale de Lausanne, Sciences et Techniques de l'Ing\'enieur, Institut des Mat\'eriaux, Laboratoire des Mat\'eriaux Magn\'etiques Nanostructur\'es et Magoniques, 1015 Lausanne, Switzerland}

\vskip 0.25cm
\date{\today}

\begin{abstract}
We report broadband spin-wave spectroscopy on kagome artificial spin ice (ASI) made of large arrays of interconnected Ni$_{80}$Fe$_{20}$ nanobars. Spectra taken in saturated and disordered states exhibit a series of resonances with characteristic magnetic field dependencies. Making use of micromagnetic simulations, we identify resonances that reflect the spin-solid-state and monopole-antimonopole pairs on Dirac strings. The latter resonances allow for the generation of highly-charged vertices in ASIs via microwave assisted switching. Our findings open further perspectives for fundamental studies on ASIs and their usage in reprogrammable magnonics.

\end{abstract}


\maketitle
Frustration\textemdash a phenomenon where competing interactions are not all satisfied at the same time\textemdash has found its consideration in different fields of science \cite{farhanprb}. Great attention was generated by pyrochlores where frustration between spins occurs due to geometric and magnetic constraints \cite{harris1997geometrical,bramwell2001spin}. Here, the microscopic spins are subject to so-called ice rules. They give rise to degenerate low-energy states and exotic phenomena such as macroscopic residual entropy \cite{ramirez1999zero}. Experimental investigations on pyrochlores are restricted in that the behavior of individual spins is inaccessible. This drawback has been overcome via artificial spin ice (ASI), i.e., an array of interacting bistable nanomagnets (macrospins) placed on a square or kagome lattice \cite{nisoli2013colloquium}. For an ASI, one can tune interactions and frustration by varying material parameters \cite{nisoli2013colloquium}. Strict spin ice behavior requires however that nanobars arranged on a square lattice possess inequivalent and fine-tuned heights  \cite{moller2006artificial,mengotti2011real}. This challenge does not exist for kagome ASI. DC magnetization studies performed on ASIs so far have generated controversies and open questions \cite{mellado2010kagome,pushp2013domain,zeissler2013non}: (I) What is the fundamental building block of the kagome ASI - the Y-shaped or bow-tie configuration? (II) Can one control the trajectory of a Dirac string created by reversed nanobars? To explore and independently test these aspects, alternative techniques are required. Applying micromagnetic simulations to a square-lattice ASI~\cite{gliga2013spectral,Gliga2015}, it was recently argued that spin dynamics reflected reversed nanomagnets and topological defects consisting of monopole-antimonopole pairs separated by Dirac strings \cite{gliga2013spectral}. It remaind unclear however how monopole-antimonopole pairs modified the dynamics of neighboring nanomagnets. For the kagome lattice, no such predictions exist either, and experimental results are lacking.\\ \indent Here, we report {\em broadband} spin-wave spectroscopy performed on a kagome ASI prepared from interconnected Ni$_{80}$Fe$_{20}$ (Py) nanobars (Fig. 1). Using micromagnetic simulations we analysed the experimental data and discovered that monopole-antimonopole pairs inside the ASI modify distinctly the internal magnetic field $\mathbf{H}_{\rm int}(x,y)$ of specific segments, giving rise to a characteristic GHz response in the crossing region of Dirac strings. These specific resonances allow one to vary locally magnetic charges within an ASI via microwave assisted switching. We also demonstrate how to discriminate between charge-ordered and spin-solid-state \cite{branford2012emerging}). Our findings are relevant for further exploration of dynamic processes in ASIs and the realization of magnonic crystals~\cite{kostylev2008partial} that are reprogrammable~\cite{topp2010making,heyderman2013artificial,krawczyk2014review,Iacocca2015}. Importantly, a kagome ASI of interconnected nanomagnets as studied here is less susceptible to a violation of the spin ice rule due to exchange interaction at the vertices \cite{qi2008direct}. It will allow one to control propagating spin waves that are exchange-dominated and key for future magnonics \cite{chumak}.\\ \indent Large arrays ($2.4~{\rm mm}\times2.4~{\rm mm}$) of kagome ASI [Fig. 1 (a)] were fabricated using nanolithography and lift-off processing of polycrystalline Py (see supplementary information). The length, width, and thickness of a Py nanobar was $l=810~$nm, $w=130~$nm, and $t=25~$nm, respectively. Broadband spin-wave spectroscopy was performed in an in-plane field $H$ using a coplanar waveguide and vector network analyzer that allowed us to measure inductively the spin-precessional motion of at least 69000 nanobars of the ASI (see supplementary information). Following Refs. \cite{gliga2013spectral}, we simulated the magnetic states and GHz dynamics in a {\em subset} of the kagome ASI using the micromagnetic code OOMMF \cite{OOMMF1} (see supplementary information).
\begin{figure}	
\includegraphics[width=0.37\textwidth]{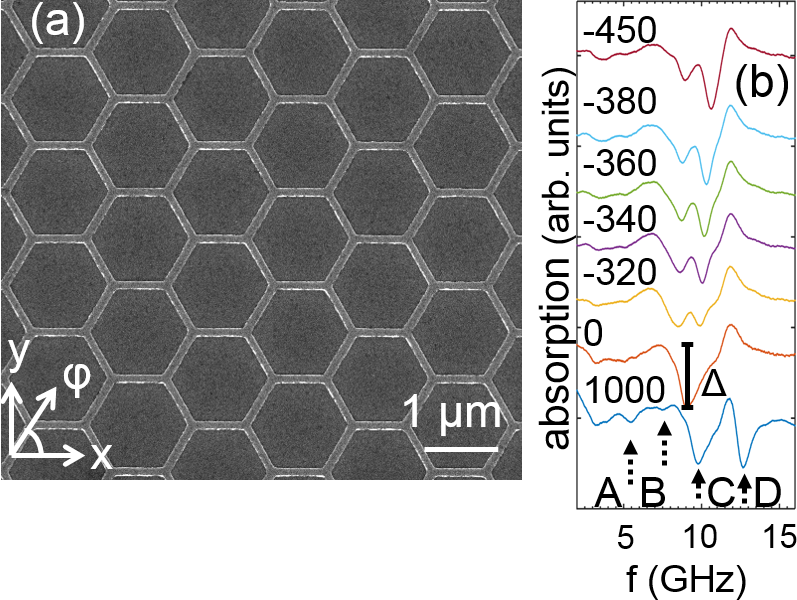}
\caption{(a) Scanning electron micrograph of interconnected Py nanobars (bright) arranged in a kagome lattice on a GaAs substrate (dark). (b) Measured spectra in the saturated and hysteretic regime for different $H$. Curves are shifted vertically for clarity. Numbers denote the field $H$ values in Oe. The positive peak at 12 GHz is a resonance contained in the subtracted reference spectrum taken at -1000 Oe with $\varphi=90^\circ$.}
\end{figure}\\ \indent
To classify different magnetic configurations, we made use of the charge model \cite{castelnovo2008magnetic,mellado2010dynamics,mengotti2011real,ladak2010direct}. We considered each Py nanobar to be a dumbbell with charges $+q$ and $-q $ at its opposing ends. Each vertex inside a kagome ASI  possessed a coordination number of 3 with a total charge $Q=\sum_{i}^{3}q_{i}$. We constructed a reproducible reference state with the following properties: (i) All nanobars had $M$ along their long axis with $+x$-component. (ii) Each vertex (excluding outer rim) obeyed a spin ice rule, i.e., $Q=+q$ or $-q$.  A monopole (antimonopole) occurred when $\Delta Q=Q_{\rm f}-Q_{\rm i}>0$ ($\Delta Q<0$), where $Q_{\rm i}$ and $Q_{\rm f}$ represented the total charge of a given vertex before and after the reversal of a neighboring segment, respectively \cite{mellado2010dynamics,mengotti2011real,ladak2010direct}.\\ \indent
Figure 1 (b) shows experimental spectra taken at different  $H$ for $\varphi=0^{\circ}$. At 1000 Oe, we label two weak ($A$ and $B$) and two strong  resonances ($C$ and $D$) (minima).  When reducing $H$ to zero, one broad resonance remains near 9 GHz with a linewidth of about 1.8 GHz. For  $H < -300  $ Oe, we observe a double-resonance structure (between about 9 and 11 GHz) that slightly shifts to larger frequency $f$ with decreasing $H$.\\ \indent Field-dependent resonance frequencies $f$ (symbols) extracted from a large series of spectra are summarized in Fig. 2 (a). For $H\geq-290~$ Oe, all modes A to D show a slope $df/dH>0$. Modes A to C are not resolved over the full field range. At -290 Oe, only one relatively broad resonance with $f=8.3$~GHz is seen. For $H=-300~$Oe, the measured spectrum changes abruptly compared to $H=-290~$Oe, and two branches evolve: (i) at 9.7 GHz, a high-frequency branch starts with $df/dH<0$ (red circles); its amplitude $\Delta$ (as defined in Fig. 1) increases significantly from -300 to -440~Oe as depicted in Fig. 2 (b) (red columns); (ii) at 8.5~GHz, a further low-frequency branch starts with $df/dH<0$ [blue symbols and columns in Fig. 2 (a) and (b), respectively]. The two branches (i) and (ii) represent modes D and C, respectively, at negative $H$. The abrupt changes in, both, $f$ and slopes $df/dH$ at -300 Oe indicate the onset of reversal in the ASI. The gradual variation of signal amplitudes (mode D) between -300 and -440~Oe suggests that the nanobars do not switch at one-and-the-same $H$ but undergo successive reversal.
\begin{figure}	
	\includegraphics[width=0.415\textwidth]{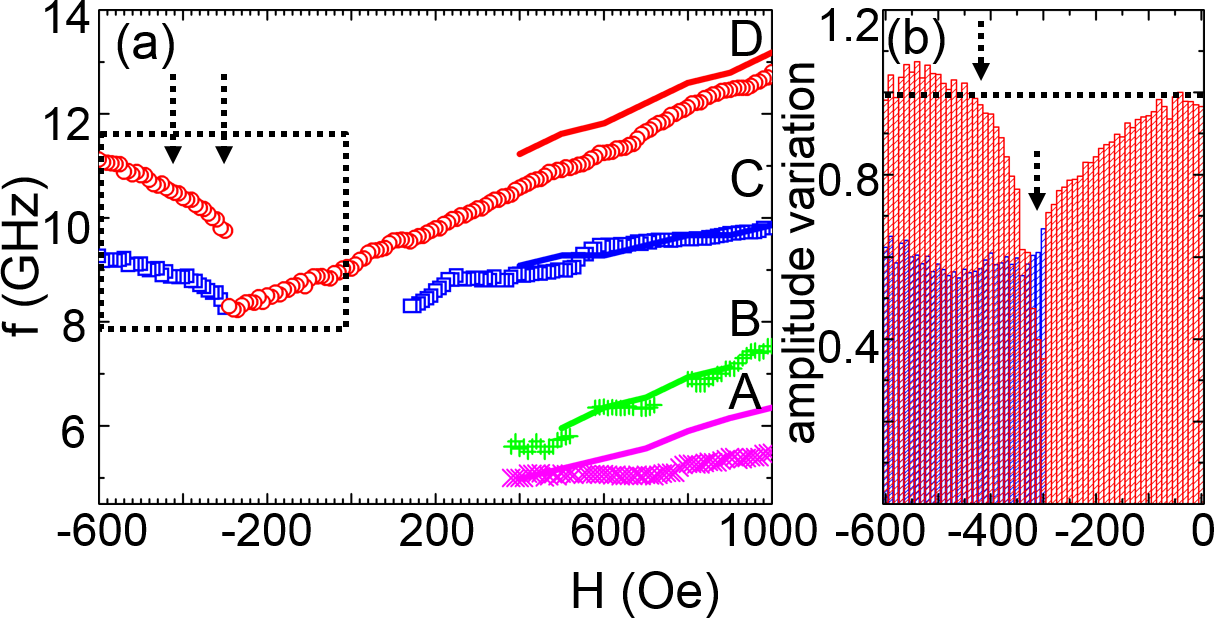}
\begin{flushleft}
\caption{(a) Experimental (symbols) and simulated (lines) resonance frequencies $f$ obtained for $H$ decreasing from 1000~Oe. Theoretical results are displayed for large $H$ [in Fig. 5 (c), we show simulated $f$ for $-295~{\rm Oe}\leq H\leq 0~$Oe]. Arrows, set at -440 and -300 Oe, border the reversal regime. The symbols size indicates the error bar in $f$. (b) Signal strength $\Delta$ of mode C resonances for $H\leq-300~$Oe (blue columns), and mode D resonances (red columns) extracted from spectra in the regime indicated by a box in (a). Values $\Delta$ are normalized to mode D at $H=-40~$Oe (horizontal line). The error margin amounts to about 0.1. Within this noise level we assume mode D to regain a saturated amplitude at -440~Oe (left arrow). The right arrow indicates $H=-300~$Oe, i.e., the onset of reversal.}
\end{flushleft}
\end{figure}
\\ \indent We now discuss the dynamic micromagnetic simulations. For large $|H|$, we extract four resonances whose field dependencies (solid lines) model reasonably well modes B to D in Fig. 2 (a).
\begin{figure}
	\includegraphics[width=0.44\textwidth]{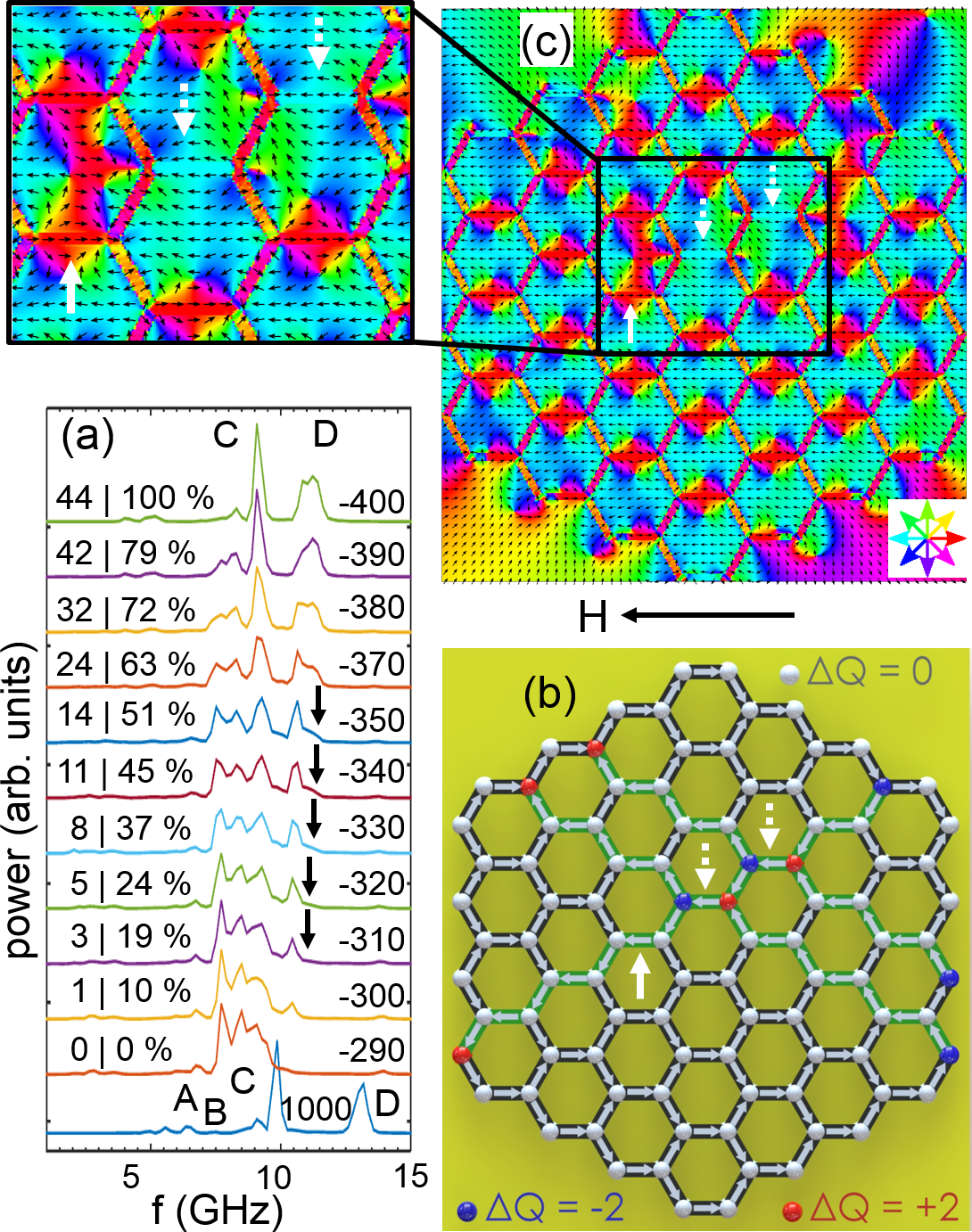}
  \begin{flushleft}
\caption{(a) Simulated power spectra for the saturated state at $H=1$~kOe, -290~Oe, and successively decreased $H$ (from bottom to top). Numbers on the left (right) indicate the number $n$ of monopole-antimonopole pairs and percentage of nanobars switched, respectively (field values in Oe). (b) Charge model representation of Dirac strings with monopole-antimonopole pairs. $\mathbf{M}$ of green (black) segments points to the left (right). Red, blue, and gray spheres represent monopoles, antimonopoles, and reference configuration, respectively. (c) Directions of the local demagnetization field $\mathbf{H}_{\rm d}(x,y)$ (black arrows and color-wheel) at $H=-320~$Oe ($\varphi=180^{\circ}$) for the ASI state shown in (b). The inset highlights the region where three Dirac strings cross. Between monopole-antimonopole pairs (downward dashed arrows) $H_{\rm d}<0$. $H_{\rm d}>0$ in remaining segments  (upward solid arrow) on the same Dirac strings.}
  \end{flushleft}
\end{figure}
Considering this agreement between experiments and simulations, we consistently label peaks with A, B, C, and D in the power spectra of Fig. 3 (a). Local power maps illustrating the spin-precessional motion (supplementary information) reveal that modes A and B reside on the edges of vertices (mode B on the outer rim). Mode C (mode D) reflects spin precession in segments enclosing an angle $\pm60 ^\circ$ ($0 ^\circ$) with $\mathbf {H}$. In the simulations, segment switching was not found down to a field value of $H=-290$~Oe. We consider the spectrum at $H=-290~$Oe [second curve from the bottom in Fig. 3 (a)] as the starting point of our following discussion. Here, a broad peak is seen extending from about 7.5 to 9.5~GHz. Its fine structure contains three closely spaced resonances at $f=7.7,8.5,~{\rm and}~9.1~$GHz. The peak at 7.7~GHz reflects spin-precessional motion in segments with $\mathbf{M}$ pointing against $\mathbf{H}$.
\\ \indent
For fields $H<-290~$Oe, we model the reversal by successively reducing $H$ and increasing the number $n$ of monopole-antimonopole pairs (Dirac strings). The assumed configurations are displayed in the supplementary information. Corresponding spectra for more and more topological defects are presented in the upper spectra of Fig. 3 (a). A single Dirac string, introduced at $H=-300~$Oe, provides an additional resonance at about 10.5 GHz. Local power maps (not shown) indicate that this mode resides in the reversed segments with $\mathbf{M}\|\mathbf{H}$ on the Dirac string. Following this resonance peak as a function of increasing $n$ (decreasing $H$), it grows in amplitude and enters the branch of mode D at negative $H$. The broad signature found between 7.5 and 9.5~GHz at -290 Oe does not vary much for $0<n<11$. For larger $n$, a pronounced peak forms that is found near 9.2~GHz at -400~Oe (uppermost curve). This peak represents mode C. Its amplitude is larger than the other peaks after more than 50~\% of the segments have been reversed.\\ \indent For $n>1$, Dirac strings can cross \cite{braun2012topological} [Figs. 3(b) and 4(a)].
\begin{figure}
	\includegraphics[width=0.38\textwidth]{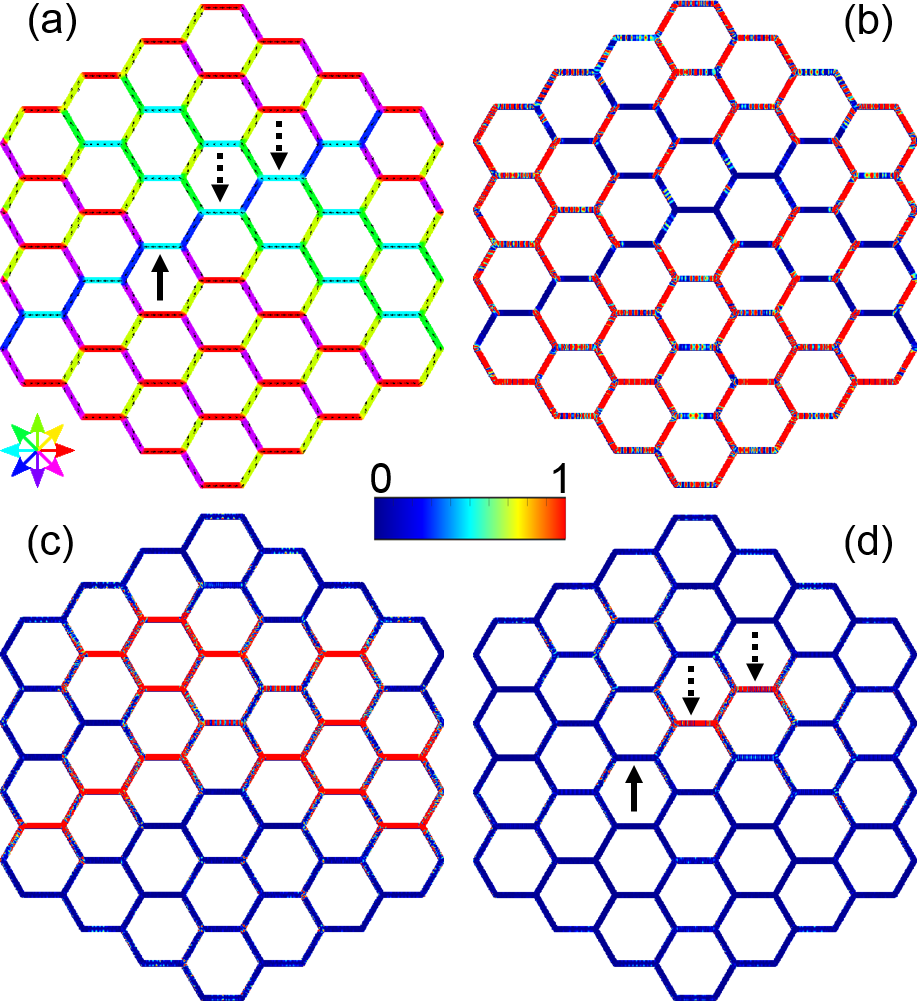}
\begin{flushleft}
\caption{(a) DC Magnetization map for the ASI state with three Dirac strings [compare Fig. 3 (b)]. Different colors represent different in-plane magnetization directions (see color-wheel). Local power maps illustrating spin-precessional amplitudes (colored legend) for (b) $f=7.7$~GHz, (c) 10.4~GHz, and (d) 11.04~GHz. Note the large amplitude in the two horizontal segments (vertical arrows) that are connected to an monopole-antimonopole pair each. White color represents non-magnetic regions.}
\end{flushleft}
\end{figure}
Figure 4 (b) to (d) displays the spin-precessional motion for $n=5$ extracted at three different eigenfrequencies $f$. The small eigenfrequency $f=7.7~$GHz [Fig. 4 (b)] represents spin-precessional motion in segments that have not been reversed at -320 Oe. For the larger $f$ of 10.4 GHz [Fig. 4 (c)], the majority of reversed segments with $\mathbf{M}\|\mathbf{H}$ precesses uniformly. However, the two segments on the Dirac strings that are surrounded by monopole-antimonopole pairs show a considerably weaker response. From the simulations we now find that crossed Dirac strings lead to a characteristic resonance feature: the mode at $f=11.04$ GHz [Fig. 4 (d)] reflects resonant spin precession between monopole-antimonopole pairs in the inner part of the ASI. These two segments [downward arrows in Fig. 3 (b)] are special in that the demagnetization field $H_{\rm d}=-34$~Oe is negative and points in the same direction as $\mathbf{H}$ [see arrows in Fig. 3 (c) and supplementary information]. Note that the sum of $H_{\rm d}$ and $H$ enters the internal field $H_{\rm int}$ and the equation of motion for the spins~\cite{Gurevich96}. Comparing Figs. 4 (c) and (d), the monopole-antimonopole pairs are thus found to locally modify $H_{\rm int}$ and increase $f$ in the crossing region of Dirac strings. The amplitude of the corresponding resonance peak in Fig. 3 (a) is a measure of number $n$ of monopole-antimonopole pairs.\\ \indent Consistent with the calculated spectra of Fig. 3 (a), we encounter a very broad resonance feature around about 8.5 GHz in the experimental data for $-290~$Oe. Also the gradual amplitude increase, seen for mode D between -300 Oe and -440 Oe in the simulations, agrees with the experiments [Fig. 2 (b)]. We have not yet resolved the specific resonance marked by black arrows in Fig. 3 (a) in our experiments, which may be due to inhomogeneous broadening when monitoring  69000 nanobars of the ASI at the same time. To evidence this mode of crossed Dirac strings, a miniaturized microwave antenna and inelastic light scattering or the magneto-optical Kerr effect might be applied \cite{Perzlmaier2005}. Using focused laser light and microwave irradiation at the relevant frequency $f$, the latter techniques provide a high spatial resolution such that the resonating segments as illustrated in Fig. 4 (d) could be identified inside an ASI.
\\ \indent We will now go one step further and discuss the spin dynamics of  states with a different degree of ordering. A kagome ASI exhibits a rich phase diagram starting from zero macroscopic correlation among  spins (gas like Ising paramagnet) to nearest-neighbor order (Ice-I phase), near-neighbour (Ice-II) and the infinite-order (spin-solid-state) ground state that consists of segments exhibiting either clockwise or anticlockwise head-to-tail magnetic moment arrangements [Fig. 5 (a)] \cite{moller2009magnetic,branford2012emerging}. Investigation of these states is important in various areas such as neural networks \cite{branford2012emerging}.
\begin{figure}
	\includegraphics[width=0.44\textwidth]{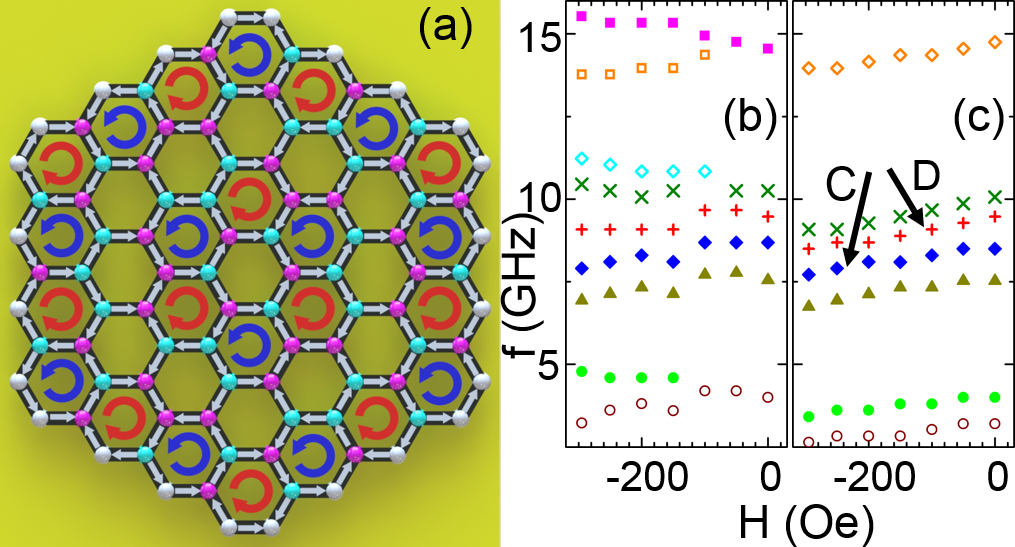}
 \begin{flushleft}
\caption{(a) Illustration of the spin-solid-state in a kagome ASI. Magenta, cyan, and gray colored spheres represent charges $+q$, $-q$, and $0$, respectively. Field dependencies of resonance frequencies (symbols) of (b) spin-solid and (c)  saturated state. We indicate branches that evolve into modes C and D at large positive $H$. The field direction is along $ \varphi=180^{\circ}$.}
 \end{flushleft}
\end{figure}
Based on our simulations, we now outline how to identify the spin-solid state [Fig. 5 (b)] comparing it with the remanent state [Fig. 5 (c)]. Spectra of both states are found to contain similar eigenfrequencies $f$ at $H=0$. Here, a clear distinction is not possible. However, applying a non-zero but small $H$ at $\varphi=180^{\circ}$ allows one to discriminate between these two states. Considering $-300~{\rm Oe}<H<0~$Oe, one avoids reversal of segments. Spectra computed for the spin-solid-state contain three groups of modes [Fig. 5 (b)]. In each group, branches coexist with, both, $df/dH>0$ and $df/dH<0$. This is strikingly different from the reference state after saturation [Fig. 5 (c)]. It supports branches with $df/dH>0$ only, as we confirmed experimentally in Fig. 2 (a). Vortex-like arrangements thus provoke additional branches with $df/dH<0$ indicating the building blocks of a spin-solid-state. The CPW-based technique is thus powerful when searching experimentally for the low-energy spin-solid-state. We expect temperature-dependent spectroscopy \cite{Schwarze2015} to be sensitive to phase transitions in ASI \cite{branford2012emerging} as well.  
\\ \indent So far, global magnetic fields were typically considered to manipulate topological defects and thereby explore the path selectivity of evolving Dirac strings~\cite{walton2015limitations,mellado2010kagome,branford2012emerging}. Our results offer a novel method to modify topological defects deterministically. By applying a microwave signal of large amplitude, one can exploit non-linear spin dynamics and microwave assisted switching \cite{nembach2007,pod2007,woltersdorf2007microwave} to reverse a selected subgroup of segments. Fine-tuning the microwave frequency to the relevant resonance frequency $f$, subgroups of the same kind are reproducibly switched in successive experiments when exploring their role as topological defects. Reversing segments embedded between monopole-antimonopole pairs [Fig. 4 (d)], one creates vertices with charges of $Q=\pm3q$. Some studies reported $Q=\pm3q$ to appear in ASIs \cite{ladak2010direct}, but later were attributed to structural disorder \cite{mellado2010dynamics}. Microwave assisted switching would create $Q=\pm3q$ intentionally inside an ideal ASI. The technique thus allows one to study whether $Q=\pm3q$ is stable or leads to an avalanche. Our work addresses also the question whether the bow-tie \cite{mellado2010dynamics} or Y-shaped configuration of nanobars \cite{pushp2013domain} forms the basic building block of a kagome ASI. Our analysis suggests that the full bow-tie subgroup rules the eigenfrequency of a nanobar and not the Y-shaped configuration.
\\ \indent In summary, we investigated kagome ASI using broadband spin-wave spectroscopy in both the saturated state and reversal regime. Based on  simulations, we explained the spectra considering topological defects. As spin waves are sensitive to small variations in the internal field, they are found to provide direct information about the occurrence of monopole-antimonopole pairs for Dirac strings that cross. The spin waves allow one to create highly charged vertices and manipulation of ASIs via microwave assisted switching. The detailed understanding of spin dynamics paves the way for reprogrammable magnonics based on ASI.
\\ \indent The research was supported by the German Excellence Cluster Nanosystems Initiative Munich II (NIM II) and the Transregio TRR80 'From electronic correlations to functionality' via the DFG.
%

\end{document}